\documentclass[12pt,aos]{article}
\usepackage{mathrsfs}
\usepackage{epsf}
\usepackage{amsthm,amsmath,amssymb,graphicx,graphics,epsfig}
\usepackage{lineno}
\usepackage{hyperref}
\usepackage[authoryear,round]{natbib}
\usepackage{longtable}
\usepackage{float}
\usepackage{multirow}
\usepackage{graphicx,graphics,epsfig}
\usepackage{cite}
\usepackage[graphicx]{realboxes}
\usepackage{rotating}
\def\E{{E}}

\def\sgn{\mbox{sgn}}

\def\bSig\mathbf{\Sigma}

\newcommand{\Rmnum}[1]{\uppercase\expandafter{\romannumeral #1}}

\begin{document}
\baselineskip 8mm \setcounter{page}{0} \thispagestyle{empty}
\begin{center}
{\Large \bf Robust penalized empirical likelihood in high dimensional longitudinal data analysis} \vspace{3mm}

Jiaqi Li$^{1}$ and Liya Fu$^{*}$  \\
{\it $^1$School of Mathematics and Statistics, Xi'an Jiaotong
University, China }\\
$^1${\it Email: lijq0305@stu.xjtu.edu.cn}\\
$^*${\it Email: fuliya@mail.xjtu.edu.cn}

\end{center}

\noindent{\bf Summary.}
As an effective nonparametric method, empirical likelihood (EL) is appealing in combining estimating equations flexibly and
adaptively for incorporating data information. To select important variables and estimating equations in a sparse high-dimensional model,
a penalized EL method based on robust estimating functions is proposed
for regularizing the regression parameters and the associated Lagrange multipliers simultaneously,
which allows the dimensionalities of both regression parameters and estimating equations to grow exponentially with the sample size.
A first inspection on the robustness of estimating equations contributing to the estimating equations selection and variable selection
is discussed from both theoretical perspective and intuitive simulation results in this paper.
The proposed method can improve the robustness and effectiveness when the data have underlying outliers or heavy tails in the response variables and/or covariates.
The robustness of the estimator is measured via the bounded influence function, and
the oracle properties are also established under some regularity conditions. Extensive simulation studies and  yeast cell data are used to evaluate the performance of the proposed
method. The numerical results reveal that the robustness of sparse estimating equations selection fundamentally enhances variable selection accuracy
when the data have heavy tails and/or include underlying outliers.


\newpage

\label{firstpage}
\section{Introduction}
\label{intro}
The variables in many scientific experiments are repeatedly recorded over time.
Scholars use these data to study how those variables of interest depend on certain characteristics of the observed subjects, and such data is called longitudinal data.
The characteristic of longitudinal data is that the measurements obtained from the same subject at different time points are correlated,
but those from different subjects are independent of each other, that is, the longitudinal data enjoys both intra-group correlation and
inter-group independence, while the true correlation structure of the data is usually unknown.
\citet{liang86} proposed the novel generalized estimating equations (GEE) method for marginal models, which suggested assigning a working matrix to replace the real
correlation matrix. They proved that whether the correlation matrix is identified correctly or not, the estimators obtained by the GEE method are consistent.
However, when the correlation matrix is misspecified, the efficiency of estimators could be reduced.
To avoid estimating the correlation parameter in a working correlation matrix,
\citet{qu00} proposed the quadratic inference functions (QIF) method,
which is more efficient than the GEE method under the same misspecification.

\citet{tang11} applied the empirical likelihood (EL) \citep{owen88} to incorporate the conditional mean model to account for
within-subject correlations via the QIF for
quantile regression in longitudinal data analysis. Compared with the traditional GEE method, the EL method is more efficient because the sandwich method usually
underestimates the variance of the GEE estimators \citep{paul15}. \citet{li13} combined QIF with empirical likelihood and studied the generalized linear models for longitudinal data.
Moreover, \citet{tian14} established the large sample property of the maximum empirical likelihood estimators in the generalized linear model.

In the past decade, variable selection has attracted more and more attention from researchers.
To select important variables in high-dimensional data, \citet{leng12} proposed a penalized empirical likelihood (PEL) approach for parameter estimation and
variable selection with the diverging dimension of covariates ($p$) and growing dimensional generalized estimating equation. More generally, \citet{tan19} extended the PEL method to generalized
linear models and established the oracle property of PEL estimators and the asymptotic distribution of the penalized empirical likelihood ratio test statistic
with fixed dimension $p$. Nevertheless, the above-mentioned procedures can only tackle the problem with $p$ no larger than the sample size $n$.
\citet{Chang18b} proposed a novel penalized empirical likelihood method (NPEL) by applying two penalty functions to regularize the model parameters and the associated Lagrange
multipliers respectively, allowing both the dimensionalities of model parameters and estimating equations to grow exponentially with the sample size.
For longitudinal data, they used the estimating equations based on QIF, and considered linear regression models.
Nevertheless, their method could be sensitive to outliers.

Robustness against outliers is a fundamental issue in longitudinal studies. When data are contaminated or follow a heavy-tailed distribution,
the penalized EL is sensitive to response and/or covariates outliers. Recently, \citet{hu2020} presented an efficient and robust EL method (EREL) by combining the robust generalized
estimating equations for longitudinal data analysis, but they only focused on robust parameter estimation. Therefore in this paper, we
propose a robust penalized empirical likelihood method to select essential variables and estimating equations simultaneously in sparse longitudinal marginal models and keep robustness when
there are outliers in the response variables and/or covariates. Furthermore, we evaluate the robustness properties of the proposed estimator from the theoretical perspective, and
prove that the proposed method has oracle properties.

The rest of the paper is organized as follows:
In Section 2, we construct a uniform robust penalized empirical likelihood ratio function (RPEL) based on QIF for high-dimensional parameter
estimation and variable selection.
A modified two-layer coordinate descent algorithm is applied to optimize the objective function and obtain an optimizer.
In Section 3, the influence functions of proposed estimators are derived and
the oracle properties of estimators are established, where we allow the dimensions of both Lagrange multipliers and estimating equations to grow
exponentially with the sample size.
In Section 4, we carry out simulation studies to evaluate the performance of the proposed method.
In Section 5, we apply the proposed method to analyze a yeast cell cycle gene expression data set.
Finally, we draw some conclusions in Section 6.

\section{Robust penalized empirical likelihood }

Suppose we have measurements
$Y_{i}=(y_{i1},\cdots,y_{im_i})^{\rm T}$ collected at times $(t_{i1},\cdots, t_{im_i})$ for subject $i$ with $n$ subjects, where $i=1,\cdots, n$.
Let $X_i=(x_{i1},\cdots,x_{im_i})^{\rm T}$ be a $m_i \times p$ matrix of covariates, where $x_{ij}$ is a $p \times 1$ vector.
The observations are independent across different subjects but can be dependent within the same subject.

In this paper, we focus on the high-dimensional sparse marginal models.
Denote the mean of  $y_{ij}$ by $\mu_{i j}=E\left(y_{i j} | {x}_{i j}\right)=g\left(x_{i j}^{\top}\boldsymbol{\beta}\right)$,
where $g(\cdot)$ is the inverse of a known link function, and
the variance of $y_{ij}$ is $\operatorname{Var}\left(y_{i j} | x_{i j}\right)=\phi v\left(\mu_{i j}\right)$ with a variance function $v(\cdot)$ and a scale parameter $\phi$.
For the sparse parameter vector $\boldsymbol{\beta}=(\beta_1,\ldots,\beta_p)^{\rm T}$, let $\mathcal{A}_0=\left\{1 \leq k \leq p: \beta_{k} \neq 0\right\}$
with $s=|\mathcal{A}_0|$, where $s$ is much smaller than $p$.
Let $\mu_i=(\mu_{i1},\ldots, \mu_{im_i})^{\rm T}$, and ${A}_{i}=\phi \operatorname{diag}\left(v\left(\mu_{i 1}\right), \ldots, v\left(\mu_{i m_{i}}\right)\right)$
be a diagonal matrix.  Assume that the working covariance matrix of $Y_i$ is ${A_i}^{1/2}{R_i}(\alpha){A_i}^{1/2}$, where
$R_i(\alpha)$ is  a working correlation matrix with a correlation parameter vector $\alpha$.
To avoid estimating  $\alpha$,
\citet{qu00} constructed the
inverse of $R_{i}(\alpha)$ by a linear combination of a class of known matrices $\{M_1,\ldots,M_l\}$, that is $R_{i}^{-1}(\alpha)=\sum_{k=1}^{l} a_{k} M_{k}$
with unknown constants $a_1,\ldots,a_l$, 
and then proposed the following estimating function
$$
\bar{g}_{n}(\beta)=\frac{1}{n} \sum_{i=1}^{n} g_{i}(\beta)=\frac{1}{n}\left(\begin{array}{c}
\sum_{i=1}^{n} D_{i}^{\mathrm{T}} A_{i}^{-\frac{1}{2}} M_{1} A_{i}^{-\frac{1}{2}}\left(y_{i}-\mu_{i}\right) \\
\vdots \\
\sum_{i=1}^{n} D_{i}^{\mathrm{T}} A_{i}^{-\frac{1}{2}} M_{l} A_{i}^{-\frac{1}{2}}\left(y_{i}-\mu_{i}\right)
\end{array}\right),
$$
where ${D}_{i}=\partial {\mu}_{i} / \partial {\boldsymbol{\beta}}=\Delta_{i} X_{i}$ with
$\Delta_{i}=\operatorname{diag}\left(\dot{\mu}_{i 1}(\beta), \ldots, \dot{\mu}_{i n_{i}}(\beta)\right)$ being a diagonal matrix of the first
derivative of the mean vector $\mu_i(\beta)$ for $i=1,\ldots,n$.
The functions $g_1(\beta),g_2(\beta),\ldots,g_n(\beta)$ are essentially based on weighted least squares, and hence are sensitive to outliers and heavy-tailed data.
To enhance the efficiency of parameter estimation, we will construct a robust estimating function in the next Subsection.

\subsection{Methodology}
Considering the robustness of the estimating function,
let
\begin{equation*}
\mathbf{g}(\mathbf{X}_i ; \boldsymbol{\beta})=\left\{g_{1}(\mathbf{X}_i ; \boldsymbol{\beta}), \ldots, g_{r}(\mathbf{X}_i ; \boldsymbol{\beta})\right\}^{\mathrm{T}}
=\left(\begin{array}{c}
{D}_{i}^{\mathrm{T}} A_{i}^{-1 / 2} M_{1} h_i(\mu_i(\beta)) \\
\vdots \\
{D}_{i}^{\mathrm{T}} A_{i}^{-1 / 2} M_{l} h_i(\mu_i(\beta))
\end{array}\right),
\end{equation*}
with ${h}_{i}\left({\mu}_{i}\right)={W}_{i}\left[\psi\left({\mu}_{i}(\boldsymbol{\beta})\right)-C_{i}\left({\mu}_{i}(\boldsymbol{\beta})\right)\right]$ for a given
bounded score function $\psi(\cdot)$,
$C_{i}\left({\mu}_{i}\right)=E\left[\psi\left({\mu}_{i}(\beta)\right)\right]$, and a diagonal weight matrix
$W_{i}$ being used to downweight the effect of leverage points, among which the $j$th diagonal element is
$$
w_{i j}=w\left(x_{i j}\right)=\min \left\{1,\left\{\frac{b_{0}}{\left(x_{i j}-m_{x}\right)^{T} S_{x}^{-1}\left(x_{i j}-m_{x}\right)}\right\}^{\frac{\varphi}{2}}\right\},
$$
where $\varphi \geq 1$, and $b_0$ is the 0.95 quantile of the $\chi^2$ distribution with $p$ degrees of freedom \citep{ter05}, $m_x$ and $S_x$ are some robust estimators of the location and scale of $x_{ij}$,
such as minimum covariance determinant (MCD) \citep{Rou84}.
In this paper, we consider Huber's score function $\psi_{c}(\cdot)$ \citep{huber64}, exponential score function $\psi_{\gamma}(\cdot)$ \citep{Frie00}, and
Tukey's Biweight score function $\psi_{b}(\cdot)$ \citep{tukey74} in the form of
$$
\psi_{c}(t)=\min \{c, \max (-c, t)\},
$$
$$
\psi_{\gamma}(t)=\frac{2 t}{\gamma} \exp \left(-t^{2} / \gamma\right),
$$
and
$$
\psi_{b}(t)=\left\{\begin{array}{ll}
{t[1-\left(\frac{t}{b}\right)^{2}]^{2}} & {\text { if }|t| \leq b} \\
{0} & {\text { if }|t|>b}
\end{array}\right.
$$
respectively, where the positive tuning parameters $c$, $\gamma$ and $b$ regulate the influence of outliers.

We apply the EL method to combine the robust generalized estimating equations based on
a chosen bounded score function $\psi(\cdot)$, 
and construct a uniform robust
penalized empirical likelihood log-likelihood ratio function as follows:
\begin{equation}\label{pel}
S_n(\boldsymbol{\beta})=\sum_{i=1}^{n} \log \left\{1+\lambda^{\mathrm{T}} \mathbf{g}\left(\mathbf{X}_{i} ; \boldsymbol{\beta}\right)\right\}-n \sum_{j=1}^{r} P_{1, v}\left(\left|\lambda_{j}\right|\right)+n \sum_{k=1}^{p} P_{2, \omega}\left(\left|\beta_{k}\right|\right),
\end{equation}
where two penalty functions $P_{1, v}(\cdot)$ and $P_{2, \omega}(\cdot)$ with tuning parameters $v$ and $\omega$ are associated with a
sparse Lagrange multiplier $\lambda$ and estimator $\boldsymbol{\beta}$ respectively, and the number of estimating equations $r=lp$.

We assume the penalty functions $P_{1,v}(\cdot)$ and $P_{2,\omega}(\cdot)$ belong to the following class as considered in \citet{lv09}:
\begin{equation*}
\begin{aligned}
&\mathcal{P}=\left\{P_{\tau}(\cdot): \rho(t ; \tau) \text { is increasing in } t \in \left[0, \infty \right) \text { and has continuous derivative } \rho^{\prime}(t ; \tau)\right. \text { for }\\
&t \in(0, \infty) \text { with } \rho^{\prime}\left(0^{+} ; \tau\right) \in(0, \infty), \text { where } \rho^{\prime}\left(0^{+} ; \tau\right) \text { is independent of } \left.\tau\right\}.
\end{aligned}
\end{equation*}
Some commonly used penalty functions such as $L_1$ penalty \citep{tib96}, SCAD penalty \citep{fan01}, and MCP penalty \citep{zhang10} all belong to class $\mathcal{P}$.
In this paper, we consider the SCAD penalty for variable selection on both sparse $\boldsymbol{\beta}$ and $\lambda$,
that is:
$$
P_{L,\theta}\left(\left|\theta\right|\right)=\eta\left|\theta\right|\left\{I\left(\left|\theta\right|<\eta\right)+\frac{\left(a-\left|\theta\right| / 2 \eta\right)}{a-1} I\left(\eta<\left|\theta\right| \leq a \eta\right)+\frac{a^{2} \eta}{(a-1) 2\left|\theta\right|} I\left(\left|\eta\right| \geq a \eta\right)\right\},
$$
where $L=1,2$, $\eta=v,\omega$ and $\theta=\lambda, \beta$ for $P_{1, v}\left(\left|\lambda\right|\right)$ and $P_{2, \omega}\left(\left|\beta\right|\right)$ respectively,
with $a=3.7$ in our simulations.


For ease and stability in implementations, taking the advice of \citet{Chang18b}, we rebuild a modified version of (\ref{pel}) to obtain the penalized EL estimator $\widehat{\boldsymbol{\beta}}$ as
\begin{equation}\label{estimator}
\widehat{\boldsymbol{\beta}}=\arg \min _{\boldsymbol{\beta}} \max _{\lambda}S_n^{\star}(\boldsymbol{\beta}).
\end{equation}
Here
$$
S_n^{\star}(\boldsymbol{\beta})=\left[\sum_{i=1}^{n} \log _{\star}\left\{1+\lambda^{\mathrm{T}} \mathbf{g}\left(\mathbf{X}_{i} ; \boldsymbol{\beta}\right)\right\}-n \sum_{j=1}^{r} P_{1, v}\left(\left|\lambda_{j}\right|\right)\right.\left.+n \sum_{k=1}^{p} P_{2, \omega}\left(\left|\beta_{k}\right|\right)\right]
$$
has a twice differentiable pseudo-logarithm function $\log _{\star}(\cdot)$ with bounded support being defined as:
$$
\log _{\star}(z)=\left\{\begin{array}{ll}
\log (z) & \text { if } z \geq \epsilon \\
\log (\epsilon)-1.5+2 z / \epsilon-z^{2} /\left(2 \epsilon^{2}\right) & \text { if } z \leq \epsilon
\end{array},\right.
$$
and $\epsilon$ is chosen as $n^{-1}$ in implementations.

\subsection{Algorithm}
To compute doubly-penalized EL with high-dimensional $p$ and $r$, we adopt the modified two-layer coordinate descent algorithm proposed in \citet{Chang18b}, where
the inner layer of the algorithm solves for $\lambda$ with a given $\boldsymbol{\beta}$ by maximizing $f(\lambda ; \boldsymbol{\beta})$, which is denoted as:
\begin{equation}\label{f_lam}
f(\lambda ; \boldsymbol{\beta})=n^{-1} \sum_{i=1}^{n} \log_{\star} \left\{1+\lambda^{\mathrm{T}} \mathbf{g}\left(\mathbf{X}_{i} ; \boldsymbol{\beta}\right)\right\}-\sum_{j=1}^{r} P_{1, v}\left(\left|\lambda_{j}\right|\right),
\end{equation}
and the outer layer of the algorithm searches for the optimizer $\widehat{\boldsymbol{\beta}}$ via
minimizing $S_n^{\star}(\boldsymbol{\beta})$ with respect to $\lambda$.
Both two layers can be solved using coordinate descent by cycling and updating each of the coordinates \citep{tang14}.
The details of implementation to the procedure are stated as follows:

Step 1. Given initial estimators $\widehat{\boldsymbol{\beta}}^{(0)}$ and $\widehat{\lambda}^{(0)}$. In our simulations, we use a robust MM estimator as $\widehat{\boldsymbol{\beta}}^{(0)}$,
and each element of $\widehat{\lambda}^{(0)}$ is set as 0.


Step 2. The $(k+1)$th update for $\widehat{\lambda}_j$ and $\widehat{\boldsymbol{\beta}}_t$ is:
$$
\widehat{\lambda}_{j}^{(k+1)}=\widehat{\lambda}_{j}^{(k)}-\frac{\sum_{i=1}^{n} \log _{\star}^{\prime}\left(t_{i}^{(k)}\right) g_{j}\left(\mathbf{X}_{i} ; \boldsymbol{\beta}\right)-n P_{1, v}^{\prime}\left(\left|\widehat{\lambda}_{j}^{(k)}\right|\right)}{\sum_{i=1}^{n} \log _{\star}^{\prime \prime}\left(t_{i}^{(k)}\right)\left\{g_{j}\left(\mathbf{X}_{i} ; \boldsymbol{\beta}\right)\right\}^{2}-n P_{1,v}^{\prime \prime}\left(\left|\widehat{\lambda}_{j}^{(k)}\right|\right)},
$$
for $j=1,\ldots,r$, where $t_{i}^{(k)}=1+\mathbf{g}\left(\mathbf{X}_{i} ; \widehat{\boldsymbol{\beta}}^{(k)}\right)^{\mathrm{T}} \widehat{\lambda}^{(k)}$.
For a given $\widehat{\lambda}=(\widehat{\lambda}_1,\ldots,\widehat{\lambda}_r)^{T}$ obtained above, we have
$$
\widehat{\beta}_{t}^{(k+1)}=\widehat{\beta}_{t}^{(k)}-\frac{\sum_{i=1}^{n} \log _{\star}^{\prime}\left(s_{i}^{(k)}\right) \varpi_{i t}^{(k)}+n P_{2, \omega}^{\prime}\left(\left|\widehat{\beta}_{t}^{(k)}\right|\right)}{\sum_{i=1}^{n}\left[\log _{\star}^{\prime \prime}\left(s_{i}^{(k)}\right) \left\{\varpi_{i t}^{(k)}\right\}^2+\log _{\star}^{\prime}\left(s_{i}^{(k)}\right) z_{i t}^{(k)}\right]+n P_{2, \omega}^{\prime \prime}\left(\left|\widehat{\beta}_{t}^{(k)}\right|\right)},
$$
for $t=1,\ldots,p$,
where
$s_{i}^{(k)}=1+{\widehat{\lambda}^{\mathrm{T}} \mathbf{g}\left(\mathbf{X}_{i} ; \widehat{\boldsymbol{\beta}}^{(k)}\right)}$,
$\varpi_{i t}^{(k)}=\widehat{\lambda}^{\mathrm{T}} \partial \mathbf{g}\left\{\mathbf{X}_{i} ; \widehat{\boldsymbol{\beta}}^{(k)}\right\} / \partial \beta_{t}$,
and
$z_{i t}^{(k)}=\widehat{\lambda}^{\mathrm{T}} \partial^{2} \mathbf{g}\left(\mathbf{X}_{i} ; \widehat{\boldsymbol{\beta}}^{(k)}\right) / \partial \beta_{t}^{2}$.

Step 3. Repeat Step 2 until convergence.

Besides, we set $\widehat{\lambda}_j=0$ for $j=1,\ldots,r$, if $|\widehat{\lambda}_j|<10^{-3}$,
and $\widehat{\beta}_t=0$ for $t=1,\ldots,p$, if $|\widehat{\beta}_t|<10^{-3}$ respectively at each iteration to ensure the sparsity of estimators.

\subsection{Robustness on the regularization parameters selection criterion}

To select important variables and estimating equations, we need to choose proper regularization parameters $v$ and $\omega$ for two penalty functions in Section 2.2,
which determine the consistency of variable selection. Denote $\tau=(v,\omega)$.
We employ the BIC-type criterion \citep{wang09} to choose the tuning parameters:
\begin{equation}\label{BIC}
\mathrm{BIC}(\tau)=2 \ell \left(\widehat{\boldsymbol{\beta}}_{\tau}\right)+\left|M_{\tau}\right| C_{p} \log (n),
\end{equation}
where
$\ell \left(\widehat{\boldsymbol{\beta}}_{\tau}\right)=\sum_{i=1}^{n} \log_{\star} \left\{1+\widehat{\lambda}^{\mathrm{T}} \mathbf{g}(\mathbf{X}_{i} ;\widehat{\boldsymbol{\beta}}_{\tau})\right\}$,
$C_p=\max(1,\log(\log p))$, and $\left|M_{\tau}\right|$ denotes the cardinality of $M_{\tau}$ with $M_{\tau}=\left\{j: \widehat{\boldsymbol{\beta}}_{j}(\tau) \neq 0\right\}$.
One can select a pair of optimal parameters $(v,\omega)$ in proper range sets by minimizing (\ref{BIC}).

A further inspection on criterion (\ref{BIC}) reveals that robust remedies on the estimating equations are necessary. Since the loss function $\ell \left(\widehat{\boldsymbol{\beta}}_{\tau}\right)$
is determined by the estimate $\widehat \lambda$, which implies estimating equations selection, and estimating equations $\mathbf{g}(\mathbf{X}_{i} ;\widehat{\boldsymbol{\beta}}_{\tau})$.
It is noteworthy that when there is severe contamination on observations, some extreme values lead to non-ignored bias on estimating equations. To maximize the penalized objective function
(\ref{f_lam}) with respect to $\lambda$, it prone to penalize less on the sparsity of the Lagrange multipliers, therefore more estimating equations are incorporated for estimating
parameter $\boldsymbol{\beta}$. In addition, significant loss induce criterion (\ref{BIC}) to select variables with a minor degree of freedom, which reduces
the accuracy of variable selection.



\section{Theoretical properties}

\subsection{Influence function}
To evaluate the local robustness, the influence function was first introduced by \citet{Ham71} to measure the stability of estimators given an infinitesimal contamination.
Suppose observations $\mathbf{Z}=(z_{1},\ldots,z_{n})$ are drawn from a common distribution $F$ over the space $\mathcal{Z}$, and a loss function $L : \mathcal{Z} \times \mathbb{R}^{p} \mapsto \mathbb{R}$
link the parameter space $\Theta \in \mathbb{R}^{p}$ with the observed data. The empirical distribution $\hat{F}=\frac{1}{n} \sum_{i=1}^{n} \delta_{z_{i}}$ with the distribution probability
$\delta_{z}$ assigning mass 1 at the point $z$ and 0 elsewhere. Then the value $E_{\hat{F}}[L(Z, \theta)]$ can be an estimator
of the unknown population risk function $E_{F}[L(Z, \theta)]$. In general, our estimator can be obtained by minimizing the penalized risk
\begin{equation}
\Lambda_{\eta}(\boldsymbol{\theta}; F)=E_{F}[L(Z, \boldsymbol{\theta})]+P(\boldsymbol{\theta} ; \eta),
\end{equation}
where the penalty function $P(\cdot; \eta)$ is $P_{1, v}\left(\cdot\right)$ or $P_{2, \omega}\left(\cdot\right)$ in Section 2.1.
Denote the first derivative of loss function $L(Z, \theta)$ as the estimating function $U\left(Z,\theta\right)$.
Denote statistical function
$T_{\theta}(F)=\theta_{F, \eta}=\theta^{*}=\operatorname{argmin}_{\theta} \Lambda_{\eta}(\theta ; F)$, then the influence function of $T$ at a point $z \in \mathcal{Z}$ for a distribution
$F$ is defined as
\begin{equation}
\operatorname{IF}_{\theta}(z ; F, T)=\lim _{\varepsilon \rightarrow 0+} \frac{T_{\theta}\left(F_{\varepsilon}\right)-T_{\theta}(F)}{\varepsilon},
\end{equation}
where $F_{\varepsilon}=(1-\varepsilon) F+\varepsilon \delta_{z}$.
In this section, we derive influence functions for the estimator $\boldsymbol{\lambda}$ and $\boldsymbol{\beta}$ respectively.

For estimator $\boldsymbol{\lambda}$, the penalized risk function is:
\begin{equation}\label{risk_lam}
\Lambda_{v}(\boldsymbol{\lambda} ; F)=-\log_{\star}\left\{1+\lambda^{\mathrm{T}} \mathbf{g}\left(\mathbf{X} ; \boldsymbol{\beta}\right)\right\}+P_{1, v}\left(\left|\lambda\right|\right),
\end{equation}
and the statistical function is denoted as
$T_{\lambda}(F)=\lambda^{*}=\operatorname{argmin}_{\lambda} \Lambda_{v}(\lambda ; F)$, then the corresponding influence function can be derived as follows:
\begin{equation}\label{influence_lam}
\begin{aligned}
\operatorname{IF}_{\lambda}(z ; F, T)&=\lim _{\varepsilon \rightarrow 0+} \frac{T_{\lambda}\left(F_{\varepsilon}\right)-T_{\lambda}(F)}{\varepsilon}\\
&=-S_{\lambda}^{-1}\left[U\left(z,\lambda^{*}\right)+\nabla P_{1,v}\left(\lambda^{*}\right)\right],
\end{aligned}
\end{equation}
where $\nabla P_{1,v}\left(\lambda^{*}\right)$ is a $r$-dimensional vector with component $P_{v,j}^{\prime}\left(|\lambda_{j}^{*}|\right)\sgn\left(\lambda_j^{*}\right)$ for $j=1,\ldots,r$.
It is natural to hold that
$S_{\lambda}^{-1}=\text{diag}\left\{\left(M_{11}+P_{1,v}^{\prime \prime}\right)^{-1},\bf 0\right\}$, where $M_{11}=E_{F}\left[\dot{U}\left(z;\lambda^{*}\right)\right]$, and $P_{1,v}^{\prime \prime}(\cdot)$ is a diagonal matrix
with the diagonal elements $p_{v, j}^{\prime \prime}\left(\left|\lambda_{j}^{*}\right|\right)$ for $j=1,\ldots,|\mathcal{G}|$,
where $|\mathcal{G}|$ is the cardinality of $\mathcal{G}=\operatorname{supp}\left\{{\boldsymbol{\lambda}}\left(\boldsymbol{\beta}\right)\right\}$ for a given estimator $\boldsymbol{\beta}$.

It is noteworthy that
\begin{equation}
\begin{aligned}
U\left(z,\lambda^{*}\right)&=-\log_{\star}^{\prime}\left\{1+\lambda^{*\mathrm{T}} \mathbf{g}\left(\mathbf{X} ; \boldsymbol{\beta}\right)\right\}\\
&=\left\{\begin{array}{ll}
\frac{-\mathbf{g}\left(\mathbf{X} ; \boldsymbol{\beta}\right)}{1+\lambda^{*\mathrm{T}} \mathbf{g}\left(\mathbf{X} ; \boldsymbol{\beta}\right)} & \text { if } 1+\lambda^{*\mathrm{T}} \mathbf{g}\left(\mathbf{X} ; \boldsymbol{\beta}\right) > \epsilon \\
\left[\left(1+\lambda^{*\mathrm{T}} \mathbf{g}\left(\mathbf{X} ; \boldsymbol{\beta}\right)\right)/\epsilon^2-2/\epsilon\right]\mathbf{g}\left(\mathbf{X} ; \boldsymbol{\beta}\right) & \text { if } 1+\lambda^{*\mathrm{T}} \mathbf{g}\left(\mathbf{X} ; \boldsymbol{\beta}\right) \leq \epsilon
\end{array}\right.
\end{aligned}
\end{equation}

Since the score function $\psi(\cdot)$ in Section 2.1 is bounded, the influence function (\ref{influence_lam}) is also a bounded function, hence we can conclude that our proposed estimator
of Lagrange multipliers is robust against outliers in either the response or the covariate domain.

Similarly, for estimator $\boldsymbol{\beta}$, the penalized risk function is:
\begin{equation}\label{risk_beta}
\Lambda_{\omega}(\boldsymbol{\beta} ; F)=\log_{\star}\left\{1+\lambda^{*\mathrm{T}} \mathbf{g}\left(\mathbf{X} ; \boldsymbol{\beta}\right)\right\}-P_{1, v}\left(\left|\lambda^{*}\right|\right)+P_{2, \omega}\left(\left|\boldsymbol{\beta}\right|\right),
\end{equation}
given the optimal estimator $\lambda^{*}$ as discussed before.
and the statistical function is denoted as
$T_{\beta}(F)=\boldsymbol{\beta}^{*}=\operatorname{argmin}_{\boldsymbol{\beta}} \Lambda_{\omega}(\boldsymbol{\beta} ; F)$, then the corresponding influence function can be derived as follows:
\begin{equation}\label{influence_beta}
\begin{aligned}
\operatorname{IF}_{\beta}(z ; F, T)&=\lim _{\varepsilon \rightarrow 0+} \frac{T_{\beta}\left(F_{\varepsilon}\right)-T_{\beta}(F)}{\varepsilon}\\
&=-S_{\beta}^{-1}\left[U\left(z,\beta^{*}\right)+\nabla P_{2,\omega}\left(\beta^{*}\right)\right],
\end{aligned}
\end{equation}
where $\nabla P_{2,\omega}\left(\beta^{*}\right)$ is a $p$-dimensional vector with component $P_{\omega,j}^{\prime}\left(|\beta_{j}^{*}|\right)\sgn\left(\beta_j^{*}\right)$ for $j=1,\ldots,p$,
and
$S_{\beta}^{-1}=\text{diag}\left\{\left(B_{11}+P_{2,\omega}^{\prime \prime}\right)^{-1},\bf 0\right\}$, where $B_{11}=E_{F}\left[\dot{U}\left(z;\beta^{*}\right)\right]$, and $P_{2,\omega}^{\prime \prime}(\cdot)$ is a diagonal matrix
with diagonal elements $p_{\omega, j}^{\prime \prime}\left(\left|\beta_{j}^{*}\right|\right)$ for $j=1,\ldots,s$.
It is noteworthy that
\begin{equation}
\begin{aligned}
U\left(z,\boldsymbol{\beta}^{*}\right)&=-\log_{\star}^{\prime}\left\{1+\lambda^{*\mathrm{T}} \mathbf{g}\left(\mathbf{X} ; \boldsymbol{\beta}^{*}\right)\right\}\\
&=\left\{\begin{array}{ll}
\frac{-\lambda^{*\mathrm{T}} \partial\mathbf{g}\left(\mathbf{X} ; \boldsymbol{\beta}^{*}\right)/\partial\boldsymbol{\beta} }{1+\lambda^{*\mathrm{T}} \mathbf{g}\left(\mathbf{X} ; \boldsymbol{\beta}^{*}\right)} & \text { if } 1+\lambda^{*\mathrm{T}} \mathbf{g}\left(\mathbf{X} ; \boldsymbol{\beta}^{*}\right) > \epsilon \\
T\left(\lambda^{*};\boldsymbol{\beta}^{*}\right) & \text { if } 1+\lambda^{*\mathrm{T}} \mathbf{g}\left(\mathbf{X} ; \boldsymbol{\beta}^{*}\right) \leq \epsilon
\end{array}\right.
\end{aligned}
\end{equation}
where $T\left(\lambda^{*};\boldsymbol{\beta}^{*}\right)=\left[\left(1+\lambda^{*\mathrm{T}} \mathbf{g}\left(\mathbf{X} ; \boldsymbol{\beta}^{*}\right)\right)/\epsilon^2-2/\epsilon\right]\lambda^{*\mathrm{T}} \partial\mathbf{g}\left(\mathbf{X} ; \boldsymbol{\beta}^{*}\right)/\partial\boldsymbol{\beta}$.

For the proposed method RPEL, it holds that
\begin{equation}
\partial\mathbf{g}\left(\mathbf{X}_i ; \boldsymbol{\beta}\right)/\partial\boldsymbol{\beta}=\left(\begin{array}{c}
D_{i}^{\mathrm{T}} A_{i}^{-1 / 2} M_{1} W_i \dot{\psi}(\mu_i)D_i \\
\vdots \\
D_{i}^{\mathrm{T}} A_{i}^{-1 / 2} M_{l} W_i \dot{\psi}(\mu_i)D_i
\end{array}\right)
\end{equation}
for $i$-th observation.
Since the derivative of $\psi(\cdot)$ is bounded with a small positive $\gamma$ for exponential score function, any $c>0$ for Huber's score function, and
any $b>0$ for Tukey's Biweight score function, the influence function (\ref{influence_beta}) is also a bounded function, hence we can conclude that our proposed estimator
of $\boldsymbol{\beta}$ enjoys robustness.

\subsection{Asymptotic properties}

Different from \citet{Chang18b}, in this Section, we will establish large sample properties of the proposed estimator when there are outliers. We focus on the
``large $p$, small $n$'' framework, which allows both $p$ and $r$ to grow exponentially with $n$. The conditions and detailed proofs are presented in the Appendix of supplemental material.

Let $\boldsymbol{\beta}_0=(\beta_{01},\ldots,\beta_{0p})^{\rm T}$ be the true value of a $p$-dimensional parameter vector
$\boldsymbol{\beta}$ with support $\boldsymbol{\Theta}$,
and $\mathcal{A}^{c}_{0}=\left\{1 \leq k \leq p: \beta_{0k} = 0\right\}$. Thus we have $\boldsymbol{\beta}_{0}=\left(\boldsymbol{\beta}_{0, \mathcal{A}_0}^{\mathrm{T}}, \boldsymbol{\beta}_{0, \mathcal{A}_0^{\mathrm{c}}}^{\mathrm{T}}\right)^{\mathrm{T}}$, where $\boldsymbol{\beta}_{0, \mathcal{A}_0} \in \mathbb{R}^{s}$ is an active (nonzero) coefficient set,
and $\boldsymbol{\beta}_{0, \mathcal{A}_0^{c}}=\mathbf{0} \in \mathbb{R}^{p-s}$.
Similarly, let the active set $\mathcal{A}=\left\{j: \widehat{\beta}_{j} \neq 0\right\}$ denote the set of indices of nonzero estimated coefficients.
Without loss of generality, we give some remarks for simplicity.
Define $\mathcal{M}_{\boldsymbol{\beta}}=\left\{1 \leq j \leq r:\left|\bar{g}_{j}(\boldsymbol{\beta})\right| \geq v \rho_{1}^{\prime}\left(0^{+}\right)\right\}$
for any $\boldsymbol{\beta} \in \Theta$, where $\rho_{1}(t ; v)=v^{-1} P_{1, v}(t)$, and
$\overline{\mathbf{g}}_{\mathcal{A}}(\boldsymbol{\beta})=n^{-1} \sum_{i=1}^{n} \mathbf{g}_{\mathcal{A}}\left(\mathbf{X}_{i} ; \boldsymbol{\beta}\right)$.
Let
$\mathbf{V}_{\mathcal{A}}(\boldsymbol{\beta})=\E\left\{\mathbf{g}_{\mathcal{A}}\left(\mathbf{X}_{i} ; \boldsymbol{\beta}\right) \mathbf{g}_{\mathcal{A}}\left(\mathbf{X}_{i} ; \boldsymbol{\beta}\right)^{\mathrm{T}}\right\}$.
Define
$$
\ell_{n}=\max _{\left\{\boldsymbol{\beta} \in \Theta:\left|\boldsymbol{\beta}_{\mathcal{A}}-\boldsymbol{\beta}_{0, \mathcal{A}_0}\right|_{\infty} \leq c_{n}, \boldsymbol{\beta}_{\mathcal{A}^{c}=0}\right\}}\left|\mathcal{M}_{\boldsymbol{\beta}}\right|
$$
for some $c_n\rightarrow 0$ satisfying $b_{n}^{1 /(2 \xi)} c_{n}^{-1} \rightarrow 0$ for a uniform constant $\xi>0$, where
$b_{n}=\max \left\{a_{n}, v^{2}\right\}$ with $a_{n}=\sum_{k=1}^{p} P_{2, \omega}\left(\left|\beta_{0k}\right|\right)$.
Under some general conditions, we present the oracle properties of the proposed estimator.

{\bf Theorem 1}
Let $P_{1, v}(\cdot), P_{2, \omega}(\cdot) \in \mathcal{P}$ and $P_{1, v}(\cdot)$ be a convex function with bounded second derivative around 0.
Let $\kappa_{n}=\max \left\{\ell_{n}^{1 / 2} n^{-1 / 2}, s^{1 / 2} \chi_{n}^{1 / 2} b_{n}^{1 /(4 \xi)}\right\}$.
Assume $\max \{\log r, \log p\}=o(n)$ with $\log r=o\left(n^{1 / 3}\right)$, $s^{2} \ell_{n}b_{n}^{1 / \xi}=o(1)$,
$\ell_{n}^{2} n^{-1}\log r=o(1)$, $\max \left\{b_{n}, \ell_{n} \kappa_{n}^{2}\right\}=o\left(n^{-2 / \zeta}\right)$,
$\ell_{n}^{1 / 2} \kappa_{n}=o(v)$, and
$\ell_{n}^{1 / 2} \max \left\{\ell_{n} v, s^{1 / 2} \chi_{n}^{1 / 2} b_{n}^{1 /(4 \xi)}\right\}=o(\omega)$.
There exist $\zeta>4$ and $\chi_{n} \rightarrow 0$, such that for a local minimizer $\widehat{\boldsymbol{\beta}}_{n} \in \boldsymbol{\Theta}$ for
(\ref{estimator}), it holds that
$$
\left|\widehat{\boldsymbol{\beta}}_{n,\mathcal{A}}-\boldsymbol{\beta}_{0, \mathcal{A}_0}\right|_{\infty}=O_{p}\left(b_{n}^{1 /(2 \xi)}\right).
$$

Theorem 1 implies the convergence rate of our estimator is $O_{p}\left(b_{n}^{1 /(2 \xi)}\right)$. According to \citet{Chang18b}, under some additional conditions,
such a rate can be improved as $O_p(v)$.

{\bf Theorem 2}
Suppose conditions in Theorem 1 hold. In addition, assume
$b_{n}=o\left(n^{-2 / \zeta}\right)$, $n s \chi_{n}^{2}=o(1)$,
$\ell_{n}^{2} (\log r) \max \left\{s^{2}\left(1+s \right) b_{n}^{1 / \xi}, n^{-1}\left(s+\ell_{n}\right) \log r\right\}=o(1)$,
$n \ell_{n} \kappa_{n}^{4} \max \left\{s, n^{2 / \zeta}\right\}=o(1)$, and
$n \ell_{n} s^{2} \max \left\{\ell_{n}^{2} v^{4}, s^{2} \chi_{n}^{2} b_{n}^{1 / \xi}\right\}=o(1)$.
As $n \rightarrow \infty$, we have
	
(1) Variable selection consistency, $P\left(\widehat{\boldsymbol{\beta}}_{n,\mathcal{A}^{c}}=\mathbf{0}\right) \rightarrow 1$, and
	
(2) Asymptotic normality:
	for any $\boldsymbol{\alpha} \in \mathbb{R}^{s} $, it holds that
	$$
	n^{1 / 2} \boldsymbol{\alpha}^{\mathrm{T}} \widehat{\mathbf{J}}_{\mathcal{G}_{n}}^{1 / 2}\left(\widehat{\boldsymbol{\beta}}_{n,\mathcal{A}}-\boldsymbol{\beta}_{0, \mathcal{A}_0}-\widehat{\boldsymbol{\psi}}_{\mathcal{G}_{n}}\right) \stackrel{d}{\rightarrow} N(0,1),
	$$
	where
	\begin{eqnarray*}
		\widehat{\mathbf{J}}_{\mathcal{G}_{n}}&=&\left\{\nabla_{\boldsymbol{\beta}_{\mathcal{A}}} \overline{\mathbf{g}}_{\mathcal{G}_{n}}\left(\widehat{\boldsymbol{\beta}}_{n}\right)\right\}^{\mathrm{T}} \widehat{\mathbf{V}}_{\mathcal{G}_{n}}^{-1}\left(\widehat{\boldsymbol{\beta}}_{n}\right)\left\{\nabla_{\boldsymbol{\beta}_{\mathcal{A}}} \overline{\mathbf{g}}_{\mathcal{G}_{n}}\left(\widehat{\boldsymbol{\beta}}_{n}\right)\right\}, \\
		\widehat{\boldsymbol{\psi}}_{\mathcal{G}_{n}}&=&\widehat{\mathbf{J}}_{\mathcal{G}_{n}}^{-1}\left\{\nabla_{\boldsymbol{\beta}_{\mathcal{A}}} \overline{\mathbf{g}}_{\mathcal{G}_{n}}\left(\widehat{\boldsymbol{\beta}}_{n}\right)\right\}^{\mathrm{T}} \widehat{\mathbf{V}}_{\mathcal{G}_{n}}^{-1}\left(\widehat{\boldsymbol{\beta}}_{n}\right)\left\{\frac{1}{n} \sum_{i=1}^{n} \frac{\mathbf{g}_{\mathcal{G}_{n}}\left(\mathbf{X}_{i} ; \widehat{\boldsymbol{\beta}}_{n}\right)}{1+\widehat{\boldsymbol{\lambda}}\left(\widehat{\boldsymbol{\beta}}_{n}\right)^{\mathrm{T}} \mathbf{g}\left(\mathbf{X}_{i} ; \widehat{\boldsymbol{\beta}}_{n}\right)}\right\},
	\end{eqnarray*}
	with $\mathcal{G}_{n}=\operatorname{supp}\left\{\widehat{\boldsymbol{\lambda}}\left(\widehat{\boldsymbol{\beta}}_{n}\right)\right\}$.

Theorem 2 establishes the oracle property of our estimator. As \citet{Chang18b} suggested, the consistency can be satisfied
by choosing $v=o\left(\min \left\{s^{-\xi} \ell_{n}^{-\xi / 2}, n^{-1 / \zeta}\right\}\right)$ and
$\omega=o\left(\min \left\{s^{-2 \xi-1} \ell_{n}^{-\xi}, s^{-1} n^{-2 / \zeta}\right\}\right)$.
The limiting distribution for $\widehat{\boldsymbol{\beta}}_{n,\mathcal{A}}$ is established provided that
$$v=o\left(\min \left\{n^{-1 / \zeta}, s^{-5 \xi / 2}(\log r)^{-\xi / 2}, n^{-1 / 4} s^{-5 / 4}\right\}\right)$$
and
$
\omega=o\left(\min \left\{n^{-2 / \zeta} s^{-1}, s^{-5 \xi-1}(\log r)^{-\xi}\right\}\right).
$

In the high-dimensional marginal model, the sparsity of the local maximizer $\widehat{\boldsymbol{\lambda}}\left(\boldsymbol{\beta}_{n}\right)=\left(\widehat{\lambda}_{n, 1}, \ldots, \widehat{\lambda}_{n, r}\right)^{\mathrm{T}}$ for $f(\boldsymbol{\lambda},\boldsymbol{\beta}_{n})$ has been established in Proposition 3 in \citet{Chang18b}, which implies that
when $\boldsymbol{\beta}_{n}$ is approaching $\boldsymbol{\beta}_{0}$, the sparse $\boldsymbol{\lambda}$ effectively conducts
a moments selection by choosing the estimating functions in a way that $\bar{g}_{j}(\boldsymbol{\beta})$ has a large absolute deviation from 0.
In summary, we have established the oracle properties for the sparse robust estimator (\ref{estimator}), including consistency in estimating nonzero components and
identifying zero components, and asymptotic normality for the estimator of the nonzero components.

\section{Simulation studies}
We present simulation studies to investigate the numerical performance of the proposed robust penalized empirical likelihood estimators, where
both continuous data and count data are considered.
To test the robustness of the proposed estimators (ERPEL for the exponential score function, HRPEL for Huber's score function and TRPEL for Tukey's Biweight score function),
we compare with 
the NPEL estimator proposed by \citet{Chang18b} and PEL estimator (PEL) proposed by \citet{leng12}.

For each procedure, a compound symmetry (CS) structure is assigned as the true correlation matrix of the response with the correlation coefficient $\alpha=0.7$.
We use two sets of basis matrices in fitting the models. We take $M_1$ to be the identity matrix, and the second basis matrix $M_2$ is a matrix with 0 on the diagonal and 1 elsewhere
for  CS structure, or a matrix with two main off-diagonals being 1 and 0 elsewhere for  AR(1) structure, respectively. The tuning parameter in the
score function $\psi(\cdot)$ is selected by minimizing the determination of the covariance matrix of the proposed estimator.
The true coefficient vector is $\boldsymbol{\beta}=(3,1.5,0,0,2,0,\ldots,0)$ with only three nonzero coefficients.
A total of $100$ replicates are generated from each of the considered models.

We evaluate the performance of these approaches in terms of the following terms:
the average number of correctly identified insignificant variables (C),
the average number of incorrectly identified significant variables (IC),
percent in terms of correctly identifying the true models (CF),
the number of selected estimating equations (No.EE),
the biases of estimators, the mean squared error (MSE) of estimators,
the median of model error (MME), 
and the average estimation error (AEE), which is the average of $\left\|\hat{\beta}-\beta_{0}\right\|^{2}$ over 100 simulations.

\subsection{Continuous Data}


In the first simulation study, we generate the continuous data from the following model:
\begin{eqnarray}\label{simu1}
y_{ij}=x_{ij1}\beta_{1}+x_{ij2}\beta_{2}+\cdots+x_{ijp}\beta_{p}+\epsilon_{ij}, ~j=1,\ldots,5, ~i=1,\ldots,n.
\end{eqnarray}
Covariates $x_{ij}=(x_{ij1},\ldots,x_{ijp})^{\rm T}$ follow a multivariate normal distribution with a mean of zero and the correlation between the $k$th and $l$th component of $x_{ij}$ being
$0.5^{|l-k|}$. The random error vectors $\epsilon_i=(\epsilon_{i1},\ldots,\epsilon_{i5})^{\rm T}$ are generated from a multivariate
Student's $t$-distribution with three degrees of freedom $T_3(0,R(\alpha))$. We set sample size $n=50$ and $p=100$.
To further illustrate our proposed method is robust, we take the following contaminated cases into consideration:

Case $1$: There is no contamination on the dataset.

Case $2$: We randomly add 10\% $y$-outliers following N(10,1) on $y_{ij}$.

Case $3$: We randomly add 5\% $x$-outliers on $x_{ij1}$ following a Student's t distribution with three degrees of freedom. Meanwhile, we change the response in the same way as Case $2$.

The simulation results are presented in Table \ref{tab1}. The proposed methods have the smallest AEE in all settings, which is evident
when the working correlation structure is correctly identified.
The proposed methods (ERPEL, HRPEL, and TRPEL) have much higher CF than other penalized empirical likelihood approaches, which represents superiority on
variable selection.
When there are outliers, NPEL incorporates more estimating equations with higher NO.EE, whereas incorrectly identified nonzero coefficients as zero (IC) in a more significant proportion than
our proposed robust methods, which verifies the statement discussed in Section 2.3.
In terms of estimation accuracy, the doubly-penalized methods contribute to
more consistent estimators than PEL under a high-dimensional setting with lower bias and MSE.
In addition, NPEL and our proposed methods, naturally control well on selecting estimating equations with lower No.EE than PEL.
As intuitive results, ERPEL and TRPEL perform better than HRPEL, especially when covariates have outliers, since Huber
estimator may not be robust against heavy-tailed covariates.

We also consider the case that the random error $\epsilon_i$ are generated from a multivariate normal distribution $N({\bf 0}, R(\alpha))$
under the same settingup as above simulation. The results have the same pattern as the heavy-tailed continuous data and are presented in Table \ref{tab2}.

%

%

\subsection{Count data}
In this Subsection, we consider the longitudinal count data.
The correlated Poisson responses $y_{ij}$ have marginal mean $\mu_{ij}$ satisfying
\begin{equation}
\mathrm{log}(\mu_{ij})=x_{ij1}\beta_{1}+\cdots+x_{ijp}\beta_{p},~ i=1,\ldots,50, ~j=1,\ldots,5,
\end{equation}
where $x_{ij1},\ldots,x_{ijp}$ are randomly generated in the same way as Section 4.1. The response data $y_{ij}$ are generated using a multivariate Poisson data
generator proposed by \citet{yahav11}. The true coefficients settings are the same as those in the continuous data. We consider the following contaminated scenarios:

Case $1^{*}$: There is no contamination on data sets.

Case $2^{*}$: We randomly add 10\% $y$-outliers following a $\chi^2$ distribution with three degrees of freedom.

Case $3^{*}$: We randomly choose 10\% of covariates $x_{i j 1}$ to be $x_{i j 1}+1$. Meanwhile, we add 10\% $y$-outliers
as Case $2^{*}$.

A summary of the performance measures is given in Table \ref{tab3}. From Table \ref{tab3} we see that, robust methods (ERPEL, HRPEL, and TRPEL) are superior to
those non-robust methods in terms of CF. NPEL seems to select fewer estimating equations than the proposed methods according to No.EE, whereas the outliers have a more
significant influence on NPEL than the proposed methods. PEL performs worse than other doubly-penalized methods in terms of MSE, AEE, MME, and CF.

\section{Yeast cell data analysis}
\label{s:real data analysis}
Identifying essential transcription factors (TFs) related to the cell cycle regulation is a general interest in gene expression study.
We utilize a yeast cell cycle gene expression data set from \citet{Spellman98}, which measures messenger ribonucleic acid levels every 7 min for 119 min with a total
of 18 measurements covering two cell cycle periods. In this section, we focus on a subset of the original data set available in PGEE package in R \citep{wang12},
which is the G1 stage in a yeast cell cycle
containing 283 genes observed over 4-time points. We include 96 TFs as covariates in the following analysis.
We consider the  same linear model as in \citet{wang12},
$$
y_{i j}=\beta_{0}+\beta_{1} t_{i j}+\sum_{k=1}^{96} \beta_{k} x_{i k}+\epsilon_{i j},~~ i=1,\ldots,283,~~ j=1,\ldots,4,
$$
where the response variable $y_{ij}$ is the log-transformed gene expression level of gene $i$ measured at time point $j$, and the covariates $x_{ik}$ are the matching score of the
binding probability of the $k$th transcription factor on the promoter region of the $i$th gene for $k=1,\ldots,96$, standardized to have mean zero and unit variance,
and $t_{ij}$ represents the time points.

We assign CS and AR(1) as the working correlation structure,
and apply the proposed robust penalized empirical likelihood methods (ERPEL, HRPEL, and TRPEL) to select TFs.
The number of selected TFs and estimating equations are summarized in Table \ref{tab4}, where we also compare with NPEL \citep{Chang18b} and PEL \citep{leng12}.
In addition, we define a relatively significant proportion representing the performance of containing the selected TFs, that is
the ratio of the number of commonly selected TFs to the total number of selected TFs using methods studied in this paper respectively.


According to the results presented in Table \ref{tab4}, both NPEL and the proposed methods select important TFs such as SWI4, SWI6 and MBP1, which have been proved to be significant
in the G1 stage \citep{wang08}, while PEL discards the SWI4 with CS working correlation. Obviously, PEL uses the most estimating equations, though it enjoys
a higher relatively significant proportion due to the least selected TFs. Our proposed robust methods are competitive with NPEL in terms of
TFs selection, from which, TRPEL performs best with higher relatively significant proportion and moderate scale of No.EE.

\section{Conclusions and Discussions}
In this paper, we consider the empirical likelihood method with robust estimating equations for high-dimensional variable selection in longitudinal marginal models.
Unlike penalized empirical likelihood method proposed by \citet{leng12}, the new penalized EL \citep{Chang18b} with double penalty functions allows
the dimensionalities of model parameters ($p$) and estimating equations ($r$) to grow exponentially with the sample size $n$,
and a drastic dimension reduction in the number of estimating equations can be achieved, which breaks the limitation of traditional penalized estimating equation
procedures where they are only capable to tackle the situation with fixed $p$ or diverging $p$ at some polynomial rate of $n$.
Our simulation results show that with properly chosen regularization parameters, NPEL method is more robust than PEL to some degree when data have heavy tails and/or
outliers appear in data. Nevertheless, its performances are unsatisfactory, especially on variable selection. A further inspection on the regularization parameters
selection reveals that the robustness on the estimating equations have impact on estimating Lagrange multipliers, which represents the estimating equations selection and
contributes to the efficiency and accuracy of the variable selection.

We combine NPEL method \citep{Chang18b}
with some robust functions based on Huber's score function, exponential score function, and Tukey's Biweight score function.
The robust doubly-penalized estimators enjoy superior robustness against outliers, and outperform no matter on parameter estimation or variable selection compared with NPEL, whereas
more estimating equations are incorporated than NPEL to find optimal tuning parameters for the bounded score functions. This is an acceptable trade-off because consistent
estimators are far more important.
Robust variable selection for ultrahigh-dimensional data is attractive in the biomedical area, and our proposed method can be extended to
cases where the dimension of covariates is in the exponential order of the sample size.
Moreover, the Lagrange multipliers selection in the empirical likelihood method may be utilized to select the correlation structure in longitudinal data, which
can be an interesting topic for future research.

\section*{Funding}

This research was supported by the National
Natural Science Foundation of China  (No.11871390).
The authors acknowledge  the support by the HPC platform at Xi'an Jiaotong University.
The authors are also grateful for the selfless reply from Professor T.T. Wu and Professor L. Yan.


\newpage

\begin{table}[!h]
	\scriptsize\centering \caption{Correlated continuous data for $p=100$ and $n=50$ with $\epsilon_{i}$ following a $T_3(0,R(\alpha))$ distribution: Comparison of PEL, NPEL, HRPEL, ERPEL, and TRPEL with working correlation matrices CS and AR(1) respectively.}\label{tab1}
	\begin{tabular}{crrrrcccccccccc}
		\hline			
		&&\multicolumn{2}{c}{$\beta_1$} &\multicolumn{2}{c}{$\beta_2$} &\multicolumn{2}{c}{$\beta_5$} &
		\multicolumn{2}{c}{ }&
		\multicolumn{2}{c}{No of Zeros}\\
		\cline{3-5}\cline{6-8}\cline{11-12} 			
		& Method & Bias & MSE & Bias & MSE & Bias & MSE  & AEE & MME  & C & IC &No.EE &CF\\
		\hline
		Case 1 &  \\
		\cline{1-1}
		CS       & PEL    &  0.076  & 1.321  & 0.035  & 0.551  & -0.012 & 0.895  & 1.447 & 0.093  & 96.81 & 0.68 & 83.91 & 63 \\
		& NPEL   &  0.019  & 0.552  & 0.001  & 0.743  &  0.018 & 0.161  & 0.557 & 0.119  & 96.37 & 0.08 & 18.08 & 72 \\
		& ERPEL  & -0.005  & 0.095  &-0.004  & 0.035  &  0.008 & 0.077  & 0.095 & 0.068  & 96.84 & 0.00 & 26.35 & 94 \\
		& HRPEL  & -0.027  & 0.063  & 0.019  & 0.030  &  0.050 & 0.053  & 0.062 & 0.047  & 96.91 & 0.00 & 24.32 & 93 \\
		& TRPEL  &  0.028  & 0.042  & 0.073  & 0.031  &  0.007 & 0.023  & 0.041 & 0.020  & 96.94 & 0.00 & 26.00 & 95 \\
		&\\
		AR(1)& PEL    &  0.175  & 0.160  &  0.273 & 0.535  &  0.270 & 0.200  & 0.150 & 0.145  & 96.67 & 0.15 & 41.93 & 66 \\
		& NPEL   &  0.010  & 0.217  & -0.126 & 0.360  & -0.025 & 0.126  & 0.215 & 0.066  & 96.74 & 0.05 & 18.90 & 86 \\
		& ERPEL  &  0.006  & 0.106  & -0.164 & 0.131  & -0.036 & 0.171  & 0.105 & 0.068  & 96.95 & 0.01 & 20.83 & 95 \\
		& HRPEL  &  0.051  & 0.119  & -0.079 & 0.197  &  0.020 & 0.147  & 0.124 & 0.065  & 96.93 & 0.06 & 19.22 & 91 \\
		& TRPEL  &  0.031  & 0.077  & -0.064 & 0.135  & -0.037 & 0.140  & 0.075 & 0.035  & 96.99 & 0.04 & 16.19 & 95 \\
		\hline
		Case 2 & \\
		\cline{1-1}
		CS       & PEL   & -0.072 & 1.851  & -1.500 & 2.885  & -2.000 & 4.951  & 2.312 & 1.550  & 96.35 & 1.29 & 71.90 & 9 \\
		& NPEL  & -0.013 & 0.673  & -0.130 & 0.492  &  0.033 & 0.898  & 0.680 & 0.289  & 94.66 & 0.18 & 21.35 & 46 \\
		& ERPEL &  0.009 & 0.108  &  0.037 & 0.040  & -0.041 & 0.098  & 0.107 & 0.074  & 96.59 & 0.00 & 21.58 & 82 \\
		& HRPEL & -0.012 & 0.118  & -0.062 & 0.160  & -0.009 & 0.195  & 0.117 & 0.121  & 96.50 & 0.04 & 18.86 & 81 \\
		& TRPEL &  0.021 & 0.132  & -0.015 & 0.089  &  0.028 & 0.079  & 0.131 & 0.089  & 96.38 & 0.00 & 19.70 & 84 \\
		&\\
		AR(1)& PEL   & -0.115 & 2.388  & -1.500 & 2.704  & -2.000 & 4.738  & 3.283 & 2.101  & 96.52 & 1.73 & 45.98 & 9 \\
		& NPEL  &  0.056 & 0.165  & -0.121 & 0.491  & -0.027 & 0.369  & 0.161 & 0.103  & 95.78 & 0.17 & 21.58 & 58 \\
		& ERPEL &  0.060 & 0.103  & -0.129 & 0.312  & -0.047 & 0.104  & 0.103 & 0.088  & 96.44 & 0.09 & 20.81 & 81 \\
		& HRPEL & -0.007 & 0.120  & -0.071 & 0.314  & -0.044 & 0.169  & 0.120 & 0.064  & 96.91 & 0.12 & 22.61 & 84 \\
		& TRPEL &  0.029 & 0.153  & -0.056 & 0.196  & -0.030 & 0.142  & 0.152 & 0.094  & 96.61 & 0.05 & 12.29 & 83 \\
		\hline
		Case 3 &   \\
		\cline{1-1}
		CS        & PEL  &-2.498  & 8.337  & -1.500 & 2.7910  &-2.000 & 4.949  & 5.309 & 1.535  & 96.91 & 1.81 & 65.95 & 26 \\
		& NPEL  & 0.031  & 0.492  & -0.093 & 0.6499  & 0.023 & 0.573  & 0.489 & 0.132  & 95.99 & 0.19 & 20.18 & 60 \\
		& ERPEL & 0.038  & 0.450  & -0.097 & 0.1715  & 0.047 & 0.177  & 0.448 & 0.039  & 96.84 & 0.07 & 20.24 & 90 \\
		& HRPEL & 0.068  & 0.067  & -0.023 & 0.1577  & 0.030 & 0.082  & 0.066 & 0.036  & 96.84 & 0.04 & 17.43 & 89 \\
		& TRPEL & 0.051  & 0.061  &  0.011 & 0.1830  & 0.058 & 0.058  & 0.062 & 0.021  & 97.00 & 0.06 & 16.59 & 94 \\
		&\\
		AR(1)& PEL & -3.000 & 11.154 & -1.500 & 2.773  & -2.000 & 4.914   & 5.4809 & 1.599  & 96.95 & 1.88 & 63.90 & 29 \\
		& NPEL  &  0.046 &  0.384 & -0.098 & 0.409  &  0.016 & 0.183   & 0.4008 & 0.082  & 96.48 & 0.17 & 20.05 & 68 \\
		& ERPEL & -0.080 &  0.352 & -0.019 & 0.347  &  0.018 & 0.224   & 0.3431 & 0.059  & 96.95 & 0.07 & 18.55 & 90 \\
		& HRPEL &  0.060 &  0.148 & -0.073 & 0.238  &  0.040 & 0.131   & 0.1425 & 0.081  & 96.96 & 0.08 & 14.62 & 88 \\
		& TRPEL &  0.073 &  0.115 & -0.118 & 0.201  & -0.009 & 0.197   & 0.1129 & 0.034  & 97.00 & 0.06 & 20.26 & 94 \\
		\hline
	\end{tabular}
\end{table}
\begin{table}[!h]
	\scriptsize\centering \caption{Correlated continuous data for $p=100$ and $n=50$ with $\epsilon_{i}$ following a multivariate normal distribution: Comparison of PEL, NPEL, HRPEL, ERPEL, and TRPEL with working correlation matrices CS and AR(1) respectively.}\label{tab2}
	\begin{tabular}{crrcccccccccccc}
		\hline			
		&&\multicolumn{2}{c}{$\beta_1$} &\multicolumn{2}{c}{$\beta_2$} &\multicolumn{2}{c}{$\beta_5$} &
		\multicolumn{2}{c}{ }&
		\multicolumn{2}{c}{No of Zeros}\\
		\cline{3-5}\cline{6-8}\cline{11-12}

		& Method & Bias & MSE & Bias & MSE & Bias & MSE  & AEE & MME  & C & IC &No.EE &CF\\
		\hline
		Case 1 & &\\
		\cline{1-1}
		CS    &  PEL & 0.150 & 0.080 &  0.232 & 0.107 & 0.161 & 0.072  & 0.079 & 0.129  & 96.83 & 0.00 & 87.74 & 84 \\
		&NPEL  & 0.019 & 0.156 & -0.020 & 0.088 &-0.032 & 0.079  & 0.154 & 0.041  & 96.86 & 0.02 & 18.09 & 86 \\
		&ERPEL & 0.020 & 0.047 & -0.018 & 0.021 & 0.001 & 0.029  & 0.046 & 0.023  & 96.87 & 0.00 & 26.04 & 93 \\
		&HRPEL & 0.015 & 0.045 & -0.019 & 0.066 & 0.007 & 0.056  & 0.046 & 0.027  & 96.95 & 0.02 & 22.04 & 94 \\
		&TRPEL & 0.023 & 0.032 &  0.044 & 0.108 & 0.009 & 0.061  & 0.031 & 0.012  & 96.98 & 0.03 & 19.46 & 95 \\
		&\\
		AR(1) &PEL   & 0.196 & 0.137 &  0.258 & 0.1166 &  0.249 & 0.095 & 0.109 & 0.158  & 96.82 & 0.00 & 29.80 & 87 \\
		&NPEL  & 0.016 & 0.183 & -0.072 & 0.2526 & -0.023 & 0.124 & 0.192 & 0.064  & 97.00 & 0.11 & 18.53 & 90 \\
		&ERPEL & 0.004 & 0.078 & -0.039 & 0.0415 & -0.014 & 0.083 & 0.078 & 0.020  & 96.66 & 0.00 & 20.70 & 96 \\
		&HRPEL & 0.017 & 0.057 & -0.021 & 0.0286 &  0.019 & 0.066 & 0.057 & 0.027  & 96.87 & 0.00 & 21.00 & 95 \\
		&TRPEL & 0.025 & 0.058 &  0.018 & 0.127  & -0.035 & 0.073 & 0.060 & 0.024  & 96.97 & 0.03 & 17.45 & 94 \\
		\hline
		Case 2 &\\
		\cline{1-1}
		CS   & PEL  &  0.203 & 0.128 &  0.122 & 0.430 &  0.186 & 0.113  & 0.116 & 0.158  & 96.88 & 0.16 & 83.81 & 80 \\
		&NPEL  & -0.032 & 0.612 & -0.028 & 0.536 & -0.049 & 0.371  & 0.619 & 0.084  & 96.62 & 0.08 & 27.75 & 81 \\
		&ERPEL &  0.069 & 0.055 & -0.013 & 0.187 & -0.004 & 0.066  & 0.055 & 0.029  & 96.80 & 0.06 & 17.38 & 90 \\
		&HRPEL &  0.057 & 0.054 &  0.005 & 0.221 & -0.003 & 0.173  & 0.055 & 0.042  & 96.93 & 0.10 & 18.19 & 87 \\
		&TRPEL &  0.041 & 0.070 & -0.010 & 0.111 & -0.009 & 0.054  & 0.068 & 0.038  & 96.92 & 0.03 & 17.57 & 90 \\
		&\\
		AR(1) & PEL  & 0.255 & 0.150 &  0.146 & 0.311 &  0.229 & 0.110 & 0.128 & 0.076  & 96.88 & 0.08 & 61.86 & 85 \\
		&NPEL  & 0.052 & 0.242 & -0.038 & 0.256 & -0.032 & 0.217 & 0.238 & 0.108  & 96.88 & 0.05 & 22.30 & 87 \\
		&ERPEL & 0.059 & 0.115 & -0.120 & 0.239 &  0.022 & 0.132 & 0.110 & 0.054  & 97.00 & 0.09 & 14.17 & 92\\
		&HRPEL & 0.028 & 0.120 & -0.100 & 0.188 & -0.068 & 0.094 & 0.118 & 0.060  & 96.92 & 0.05 & 19.00 & 94 \\
		&TRPEL & 0.073 & 0.148 & -0.194 & 0.268 & -0.035 & 0.098 & 0.141 & 0.094  & 96.99 & 0.07 & 12.31 & 92  \\
		\hline
		Case 3 & &\\
		\cline{1-1}
		CS &PEL   &  0.140 & 0.082 &  0.139 & 0.256 & 0.176 & 0.091  & 0.078 & 0.116  & 96.57 & 0.07 & 99.80 & 72 \\
		&NPEL  & -0.008 & 0.534 & -0.028 & 0.537 & 0.033 & 0.190  & 0.530 & 0.083  & 96.67 & 0.12 & 22.08 & 74 \\
		&ERPEL &  0.069 & 0.092 &  0.002 & 0.125 & 0.006 & 0.126  & 0.086 & 0.040  & 96.82 & 0.03 & 15.94 & 91 \\
		&HRPEL &  0.032 & 0.058 & -0.001 & 0.161 & 0.021 & 0.066  & 0.057 & 0.023  & 96.89 & 0.05 & 17.54 & 85 \\
		&TRPEL &  0.052 & 0.076 & -0.043 & 0.142 & 0.011 & 0.093  & 0.075 & 0.051  & 96.90 & 0.03 & 17.23 & 94 \\
		&\\			
		AR(1) &PEL   &  0.160 & 0.111 &  0.141 & 0.286 &  0.228 & 0.109 & 0.096 & 0.120  & 96.78 & 0.08 & 91.78 & 73 \\
		&NPEL  & -0.038 & 0.237 & -0.007 & 0.264 &  0.034 & 0.179 & 0.236 & 0.081  & 96.69 & 0.06 & 20.44 & 77 \\
		&ERPEL &  0.013 & 0.091 & -0.005 & 0.130 & -0.043 & 0.108 & 0.094 & 0.094  & 96.97 & 0.03 & 13.97 & 94 \\
		&HRPEL &  0.065 & 0.146 & -0.061 & 0.271 &  0.017 & 0.126 & 0.140 & 0.117  & 97.00 & 0.09 & 13.23 & 91 \\
		&TRPEL &  0.066 & 0.134 &  0.010 & 0.199 & -0.003 & 0.118 & 0.128 & 0.075  & 96.99 & 0.06 & 14.00 & 93 \\
		\hline
	\end{tabular}%
\end{table}

\begin{table}[!h]
	\scriptsize\centering \caption{Correlated count data for $p=100$ and $n=50$: Comparison of PEL, NPEL, HRPEL, ERPEL, and TRPEL with
		working correlation matrices CS, AR(1) respectively.}\label{tab3}
	\begin{tabular}{crrcccccccccccc}
		\hline			
		&&\multicolumn{2}{c}{$\beta_1$} &\multicolumn{2}{c}{$\beta_2$} &\multicolumn{2}{c}{$\beta_5$}&&&\multicolumn{2}{c}{No of Zeros}\\
		\cline{3-5}\cline{6-8}\cline{11-12}

		& Method & Bias & MSE & Bias & MSE & Bias & MSE & AEE & MME  & C & IC &No.EE &CF\\
		\hline
		Case $1^*$ &  \\
		\cline{1-1}
		CS    & PEL   &  0.028 & 1.295 &  0.032 & 0.357 &  0.062 & 0.608 & 1.458   & 0.118  & 96.29 & 0.48 & 111.92 & 46 \\
		& NPEL  &  0.009 & 0.119 &  0.025 & 0.053 &  0.018 & 0.071 & 0.124   & 0.006  & 96.90 & 0.00 & 7.12   & 92 \\
		& ERPEL & -0.001 & 0.089 & -0.035 & 0.043 & -0.008 & 0.066 & 0.098   & 0.007  & 96.99 & 0.00 & 25.85  & 99 \\
		& HRPEL & -0.026 & 0.069 & -0.023 & 0.016 &  0.013 & 0.042 & 0.076   & 0.008  & 96.99 & 0.00 & 42.92  & 99 \\
		& TRPEL &  0.001 & 0.084 &  0.010 & 0.030 & -0.005 & 0.052 & 0.091   & 0.005  & 97.00 & 0.00 & 32.69  & 100 \\
		&\\
		AR(1) & PEL   &  0.034  & 1.284 &  0.036 & 0.3526 &  0.027 & 0.588 & 1.459 & 0.090  & 96.20 & 0.48 & 104.34 & 43 \\
		& NPEL  &  0.007  & 0.108 &  0.026 & 0.0492 &  0.028 & 0.065 & 0.111 & 0.006  & 96.91 & 0.00 &   6.60 & 94 \\
		& ERPEL & -0.017  & 0.129 & -0.019 & 0.0590 &  0.004 & 0.070 & 0.154 & 0.008  & 97.00 & 0.00 &  23.10 & 100 \\
		& HRPEL & -0.042  & 0.084 & -0.021 & 0.0240 & -0.023 & 0.058 & 0.109 & 0.009  & 97.00 & 0.00 &  41.61 & 100 \\
		& TRPEL & -0.019  & 0.133 & -0.026 & 0.0550 & -0.009 & 0.082 & 0.162 & 0.009  & 96.99 & 0.00 &  34.15 & 99 \\
		\hline
		Case $2^{*}$ &\\
		\cline{1-1}
		CS  & PEL    & -0.034 & 1.788 & -0.039 & 0.472 & -0.011 & 0.815  & 2.301 & 0.428  & 94.80 & 0.73 & 113.35 & 25 \\
		& NPEL   & -0.023 & 0.298 & -0.006 & 0.130 &  0.007 & 0.181  & 0.330 & 0.021  & 96.49 & 0.03 & 8.35   & 74 \\
		& ERPEL  & -0.060 & 0.370 & -0.074 & 0.209 & -0.031 & 0.245  & 0.461 & 0.017  & 96.78 & 0.08 & 21.05  & 84 \\
		& HRPEL  & -0.087 & 0.507 & -0.029 & 0.261 & -0.032 & 0.283  & 0.684 & 0.020  & 96.93 & 0.13 & 36.68  & 83 \\
		& TRPEL  & -0.060 & 0.310 & -0.066 & 0.166 & -0.037 & 0.198  & 0.375 & 0.017  & 96.79 & 0.05 & 26.56  & 85 \\
		&\\
		AR(1)& PEL   & -0.007 & 1.193 & -0.014 & 0.332 &  0.013 & 0.577  & 1.398 & 0.208  & 94.99 & 0.44 & 128.77 & 27 \\
		& NPEL  & -0.027 & 0.327 & -0.010 & 0.130 &  0.007 & 0.195  & 0.372 & 0.012  & 96.57 & 0.03 & 8.14   & 79 \\
		& ERPEL & -0.057 & 0.504 & -0.032 & 0.164 & -0.034 & 0.257  & 0.645 & 0.022  & 96.85 & 0.06 & 17.58  & 86 \\
		& HRPEL & -0.118 & 0.651 & -0.085 & 0.280 & -0.047 & 0.352  & 0.969 & 0.025  & 96.93 & 0.12 & 27.12  & 84 \\
		& TRPEL & -0.057 & 0.445 & -0.069 & 0.172 & -0.022 & 0.225  & 0.543 & 0.017  & 96.79 & 0.05 & 28.41  & 84 \\
		\hline
		Case $3^{*}$ &\\
		\cline{1-1}
		CS  & PEL    & -0.009 & 1.545 & -0.040 & 0.442 & -0.013 & 0.777 & 1.889 & 0.226  & 95.61 & 0.63 & 106.15 & 36 \\
		& NPEL   &  0.002 & 0.241 & -0.040 & 0.123 & -0.019 & 0.123 & 0.255 & 0.018  & 96.81 & 0.01 & 7.13   & 91 \\
		& ERPEL  &  0.005 & 0.230 & -0.077 & 0.116 & -0.038 & 0.126 & 0.249 & 0.020  & 96.91 & 0.02 & 20.13  & 96 \\
		& HRPEL  & -0.021 & 0.118 & -0.061 & 0.032 & -0.037 & 0.075 & 0.137 & 0.012  & 96.93 & 0.00 & 32.24  & 95 \\
		& TRPEL  & -0.012 & 0.242 & -0.038 & 0.102 & -0.031 & 0.122 & 0.271 & 0.015  & 96.93 & 0.02 & 26.31  & 95 \\
		&\\
		AR(1)& PEL   &  0.027 & 1.234 & -0.029 & 0.335 &  0.013 & 0.606 & 1.402 & 0.129  & 95.43 & 0.46 & 107.17 & 33 \\
		& NPEL  & -0.005 & 0.154 & -0.035 & 0.075 & -0.002 & 0.092 & 0.160 & 0.014  & 96.76 & 0.00 & 7.27   & 92 \\
		& ERPEL & -0.009 & 0.192 & -0.067 & 0.076 & -0.047 & 0.096 & 0.214 & 0.011  & 96.93 & 0.01 & 21.43  & 93 \\
		& HRPEL & -0.023 & 0.132 & -0.048 & 0.032 & -0.040 & 0.087 & 0.155 & 0.012  & 96.92 & 0.00 & 38.11  & 95 \\
		& TRPEL & -0.021 & 0.241 & -0.046 & 0.081 & -0.044 & 0.129 & 0.272 & 0.012  & 96.91 & 0.00 & 33.81  & 94 \\
		\hline
	\end{tabular}%
\end{table}

\begin{table}[!h]
	\scriptsize\centering \caption{Number of TFs selected for the G1-stage yeast cell-cycle process, the proportion of containing relatively significant
		reported TFs (in parentheses), and
		the number of estimating equations (No.EE) selected by PEL, NPEL, ERPEL, HRPEL, and TRPEL under
		with CS, AR(1) working correlation matrices respectively.}\label{tab4}
	\begin{tabular}{l|cc|cc}
		\hline			
		\multicolumn{1}{c}{ }&
		\multicolumn{2}{|c|}{CS} &
		\multicolumn{2}{c}{AR(1)}\\
		\hline	
		Methods & TFs & No.EE & TFs & No.EE\\
		\hline
		PEL& 23(39.13\%)& 196& 24(37.50\%)& 196\\
		NPEL& 32(32.15\%) & 2 & 33(30.30\%) & 2 \\
		ERPEL& 31(32.25\%) & 7  &31(32.25\%) & 12\\
		HRPEL&33(30.30\%) & 2 &33(30.30\%) & 7 \\
		TRPEL&26(38.46\%) & 4 & 26(38.46\%) & 9 \\
		\hline
	\end{tabular}%
\end{table}

\label{lastpage}
\end{document}